\documentclass[11pt,twoside]{article}

\usepackage{asp2006}
\usepackage{epsf}
\usepackage{lscape}

\begin{document}
\title{Connectivity in the Astronomy Digital Library}
\author{G\"unther~Eichhorn, Alberto~Accomazzi,
Carolyn~S.~Grant, Edwin~A.~Henneken, Donna~M.~Thompson,
Michael~J.~Kurtz, Stephen~S.~Murray}
\affil{Harvard-Smithsonian Center for Astrophysics, 60 Garden Street,
Cambridge, MA 02138}

\begin{abstract}

The Astrophysics Data System (ADS) provides an extensive system of
links between the literature and other on-line information.  Recently,
the journals of the American Astronomical Society (AAS) and a group of NASA
data centers have collaborated to provide more links between on-line data
obtained by space missions and the on-line journals. Authors can now
specify which data sets they have used in their article.  This
information is used by the participants to provide the links between
the literature and the data.

\medskip
\noindent
The ADS is available at:

\medskip

http://ads.harvard.edu

\end{abstract}

\section{Introduction}

The Smithsonian/NASA Astrophysics Data System (ADS) provides access to
the astronomy and physics literature.  As of September 2006 it
provides a search system for almost 4.9 million records, covering all
of the astronomical literature (including planetary sciences and solar
physics) and a large part of the physics literature (including
geosciences).  The ADS has been described in detail in a series of
articles in Astronomy and Astrophysics Supplements
\citep{2000A&AS..143...41K,2000A&AS..143...61E,2000A&AS..143...85A,
2000A&AS..143..111G}.

The literature in astronomy is by now almost completely available
on-line.  All data collected by space missions, as well as many data
obtained from ground based observations are available on-line as well.
In order to facilitate astronomy research, it is desirable to connect
the literature with the data that are used in the research articles.

This article describes some aspects of the linking between the ADS and
other on-line resources.  In particular, it describes the
collaborative effort between the NASA data centers and the American
Astronomical Society's publisher, the University of Chicago Press
(UChP), to improve and extend the linking between the literature and
on-line data.

\section{Current Data in the ADS}

As of September 2006, the ADS holds almost 4.9 million records, 2/3 of
these have abstracts, the rest are titles only.  These records are
divided in four databases (astronomy, including planetary sciences and
solar physics, physics, including geosciences, arXiv e-prints, and
general sciences).  There are a total of almost 14 million links in
the ADS, about 5 million of these are links to outside resources, the
remainder are links internal to the ADS.  Internal links are links to
abstracts, scanned articles, reference and citation lists, among
others.  Internal links point to about 3.2 million abstracts, 1.7
million reference lists and 1.7 million citation lists.  External
links point to on-line journals, other library systems, and on-line
data.  Of the external links almost 325,000 point to on-line data in
various forms.  Table~\ref{linkletters} shows the list of links in the
ADS.

\begin{table*}
\caption[]{Link types in the ADS database.
}
\label{linkletters}
\begin{tabular*}{5.2in}{llp{0.544\linewidth}}
\noalign{\smallskip}
\hline
\noalign{\smallskip}

Link & Resource & Description\\
\hline
\noalign{\smallskip}
A & Abstract & Full abstract of the article.  These abstracts come
                    from different sources.\\

C & Citations & A list of articles that cite the current article.
                    This list is not necessarily complete (see `R'
                    References).\\

D & On-line Data & Links to on-line data at other data centers.\\

E & Electronic Article & Links to the on-line version of the article.
                    These on-line versions are in HTML format for
                    viewing on-screen, not for
                    printing.$^{\mathrm{a}}$\\

F & Printable Article & Links to on-line articles in PDF or
                    Postscript format for
                    printing.$^{\mathrm{a}}$\\
                     
G & GIF Images & Links to the images of scanned articles in the ADS
                    Article Service.\\

H & HEP/SPIRES & Links to the High Energy Physics digital library SPIRES.\\

I & Author Comments & Links to author supplied additional
					information (e.g. corrections, additional
	                references, links to data),\\

J & Document Delivery & Links to on-line document delivery systems at the
                    publisher/owner of the article.\\

K & Library Entries & Links to entries in various library systems.\\

L & OpenURL & Links to an OpenURL server.\\

M & Multi-media & Links to on-line multi-media information.\\

N & NED Objects & Access to lists of objects for the current article
                    in the NED database.\\

O & Associated Articles & A list of articles that are associated
                    with the current article.  These can be errata or
                    other articles in a series.\\
P & Planetary Data System & Links to data in the Planetary Data
					System.\\

R & References & A list of articles referred to in the current
                    article.  These lists are not
                    necessarily complete, they contain only references
                    to articles that are in the ADS database.\\

S & SIMBAD Objects & Access to lists of objects for the current
                    article in the SIMBAD database.\\

T & Table of Contents & Links to the list of articles in a books or
					proceedings volume.\\

U & Also-Read Articles & Links to the list of articles that were read
                    by the same people that read the current article.\\

X & arXiv e-prints & Links to the arXiv e-print of an article.\\

Z & Custom format & Link to the abstract formatted according to the
                    user's preferences.\\
\hline
\end{tabular*}
\begin{list}{}{}
\item[$^{\mathrm{a}}$]There
is generally access control at the site that serves these on-line articles
\end{list}{}{}
\end{table*}

A more detailed description of the resources in the ADS that these links
point to is provided in \cite{2000A&AS..143..111G}.

\section{External Links}

\subsection{On-line Journals}

Most journal publishers work with the ADS to facilitate linking to the
on-line journals from the ADS.  They provide us with the information
how to build these links.

Since on-line journals generally require subscriptions, users need
to authenticate themselves.  This can be done in various ways:

1. Password authentication:  If the user has an account name and
   password, authentication can be done directly by the user.

2. IP address authentication:  Some publishers provide access on an IP
   address basis.  This means that users can access these publishers
   from their computers at work.

3. Proxy servers: Some libraries have set up proxy servers that are
   authenticated with some publishers.  If access to the articles is
   done through the proxy server, the user has direct access.  This
   works from any computer, not just from a restricted range of IP
   addresses.  The user needs to authenticate with the proxy
   server and then has access to the participating publishers.  The
   ADS fully supports the use of proxy servers.  It can be configured
   so that only external links go through the proxy server.

4. OpenURL servers:  Links to an OpenURL server at a library provides
   a service similar to the proxy server system, except that all other
   links can go directly to their destination without having to go
   through a proxy server.  The ADS fully supports OpenURL linking.

The ADS has recently (in September 2006) implemented OpenURL linking.
This new capability makes use of OpenURL servers that are set up by
libraries or institutions.  It makes it much easier for users to
access subscription journals.  In the ADS Preferences section is now a
link to a page for ``OpenURL Settings''.  This page allows our users
to specify their OpenURL server.  In order to make it easier for our
users, we provide a list of servers for major institutes for selection
on this page.  If you have an OpenURL server that is not in this list,
let us know about it and we will include it.

Any library that has an OpenURL server can include a link to our setup
page on their library page, which includes their OpenURL server.
If a user of that library follows this link, s/he will have the
OpenURL address automatically filled in.

The link to this page for libraries has the following form:

\smallskip
\begin{verbatim}
http://adsabs.harvard.edu/cgi-bin/pref_set?4&
      OpenURL=serverURL&
      Icon=iconURL
\end{verbatim}
\smallskip

The parameter OpenURL specifies the URL of the OpenURL server of the
library.  The parameter Icon specifies the name of the icon that is
used for the OpenURL links.  If it is specified without a full URL (as
in the above example), the server OpenURL is pre-pended.  The parameter
names are case sensitive.

\subsection{Data Linking}

For astronomers it is frequently important to be able to access data
that were used in an article, as well as the data published in the
article (data tables, etc.).  The ADS is provided with the information
on what data are correlated with articles on a regular basis by
several data centers.  The ADS includes the links from the articles to
the on-line data.

There are two types of data that the ADS links to.  One type are data
in data services that aggregate and process information, the other are
original data collected by telescopes or spacecraft.

Aggregated data are collected by several data archives.  The most
important of these archive services are:

\smallskip
SIMBAD \citep{2000A&AS..143....9W} for
information about astronomical objects

Vizier
\citep{2000A&AS..143...23O} for data catalogs and data tables

NASA Extragalactic Database \citep{1992adass...1...47M} for
extragalactic objects

\smallskip
The distribution of the links to the aggregating data archives is by
now fully automated.  There are on the order of 220,000 links to these
services in the ADS.

The archives that hold original data also provide some information on
what data were used in articles.  This information has so far been
collected by hand by the different data centers that hold the data.

Since collecting this information is a very time consuming task, the
Astrophysics Datacenter Executive Committee (ADEC) initiated efforts
to improve the linking between journal articles and on-line data.  The
data centers, the ADS and the UChP developed a system that allows
authors to specify which data sets they have used for their article.
The system was designed to allow for automatically processing and
verifying data set identifiers specified by authors.  Following is a
brief description of this system.

\subsubsection{Data Set Identifiers}

The basis for this system is the identification of data sets.  The
data centers assign unique identifiers to each set of data.  It is up
to the data center to decide what they call a data set.  It could be
one spectrum, or one exposure, or it could be a set of exposures of
the same object in different wavelengths.  This is left completely up
to the data centers.  In some cases, data sets are defined by the
query parameters to a database query.  Some data centers also provide
the means for authors to define a collection of data sets that they
used in an article and give this collection a unique identifier.  The
main requirement for data set identifiers is that they have to be
unique and permanent.  This means that the data centers have to agree
to recognize published identifiers in perpetuity.  This is extremely
important for the long term viability of this system.

The ADEC has agreed on a format for data set identifiers that is
compatible with current International Virtual Observatory Alliance
(IVOA, \cite{2004SPIE.5493..137Q}) designs for identifiers:

\smallskip

ADS/FacilityId\#PrivateId

\smallskip
``ADS'' specifies the ADS as the managing authority for these
identifiers.  Verification and linking is done through the ADS master
verifier and link resolver.  ``FacilityId'' specifies the facility
that collected the data, and ``PrivateId'' is an identifier assigned
by the data center.  The ADEC decided that the data center should not
be part of the identifiers, since data sets can potentially move
between data centers, which would invalidate identifiers that contain
the data center.

The data centers provide the data set identifiers with each data
set that they send to their users.  These identifiers should be
prominently visible so that authors can easily find them and include
them in their manuscripts.

\subsubsection{Identifier Verification}

Once publishers receive manuscripts that contain data set identifiers,
they verify that the identifiers are correct.  During the verification
process they obtain the permanent link for identifiers that are valid.
This is done through the ADS Verifier.  The verifier can be accessed
through SOAP (Simple Object Access Protocol) or through a simple CGI
interface.  During copy-editing, the publisher sends the data set
identifiers cited in the paper to the ADS master verifier for
verification.  The master verifier contacts the relevant local
verifier at the data center that currently has the data sets for the
facility specified in the identifier.  It then returns the status of
the identifier as returned from the data center verifier, and the
permanent link to the data set if it is a valid identifier.

\subsubsection{Data Set Links}

The permanent links in the on-line journal article to data sets do not
point directly to the data center, but rather to the ADS link
resolver.  The reason for this is that data sets can move
between data centers.  The ADS is automatically kept up-to-date about
the location of all data sets and forwards data set link requests to
the data center that currently holds the data.  The link resolver
consults the current data center profiles to determine which data
center currently holds the data for the specified FacilityId, and
retrieves the current link to the specified data set from the data
center.  It then forwards the request to that address.  This assures
that the links in the on-line journals are permanent and do not have
to be changed if data sets move.

\subsubsection{Link Distribution}

In order to fully utilize the linking information, the ADS harvests
the correlation between data set identifiers and articles from the
participating publishers.  This information is then used to link from
the ADS records to the on-line data.  The ADS also makes these
correlations available to the participating data centers.  We provide
an HTTP interface for harvesting of these correlations by data
centers.  The data centers use this information to link from their
data back to the journal articles.

\section{Current Status}

The following are the data centers that currently participate in this
system:

\smallskip
Chandra X-ray Center (CXC)

High Energy Astrophysics Science Archive Research Center (HEASARC)

Infrared Science Archive (IRSA)

Legacy Archive for Microwave Background Data Analysis (LAMBDA)

Multimission Archive at Space Telescope Science Institute (MAST)

Spitzer Science Center (SSC)
\smallskip

Currently the only publisher participating is the University of
Chicago Press, publisher of the Astronomical Journal, the
Astrophysical Journal with its Letters and Supplements, and the
Publications of the Astronomical Society of the Pacific.  Currently
there are less than 100 articles with a few hundred identifiers that
used this system.  Hopefully this number will increase as more authors
become aware of this new capability.

\section{The Future}

We encourage other data centers and publishers to participate in this
service.  The ADS has software packages that facilitate the setup of
the data center registration, identifier verification and linking
services.

The requirements for a data center to participate are as follows:

1. Provide data set identifiers with the data in the data center.
   These identifiers need to be available to the users of the data.
   They need to be permanent identifiers.

2. Provide a data center profile.  This will let the ADS determine
   which data sets are at that data center.  This profile is a simple
   XML file.  It needs to be in a specific location.  The ADS will
   automatically retrieve this profile regularly to get the latest
   status of the data center.

3. Provide a data set identifier verification utility.  The ADS master
   verifier connects to this verification utility when a publisher
   tries to verify an identifier.  This utility also needs to provide
   the current links to the data.

The ADS has examples of profiles and software toolkits for a verifier
available, in order to facilitate the participation of data centers in
this system.

The requirements for a publisher to participate are as follows:

1. Provide a macro in the LaTeX macros that allows authors to specify
   data set identifiers.

2. Verify identifiers that authors specify during the editorial
   process.

3. Link to data sets from the on-line journal.  This is optional, but
   encouraged.

4. Provide the correlation between identifiers and articles for
   harvesting by the ADS.  This can be done either by providing the
   data for downloading by the ADS, or by including them in the
   regular abstract feed to the ADS.

\section{Conclusion}

The efforts to improve the linking between the literature and on-line
data have only now started.  There are already some links to data in
the AAS journals, and more are coming on-line.  As this system becomes
more widely known the number of links should increase.  Hopefully more
data centers and publishers will join this effort to increase the
coverage of this system and provide more utility to the astronomical
community.  

\section{Links}
ADEC: http://www.adccc.org/\\
ADS: http://ads.harvard.edu/\\
ADS Data Linking: http://vo.ads.harvard.edu/\\
CXC: http://asc.harvard.edu/\\
HEASARC: http://heasarc.gsfc.nasa.gov/\\
IRSA: http://irsa.ipac.caltech.edu/\\
LAMBDA: http://lambda.gsfc.nasa.gov/\\
MAST: http://archive.stsci.edu/mast.html\\
NED: http://nedwww.ipac.caltech.edu/\\
SIMBAD: http://simbad.u-strasbg.fr/\\
SSC: http://ssc.spitzer.caltech.edu/\\
Vizier: http://vizier.u-strasbg.fr/\\

\acknowledgements

The ADS is funded by NASA Grant NNG06GG68G.

\end{document}